\documentclass[prl,twocolumn,draft,amsmath]{revtex4}
\usepackage{graphics}

\begin{document}
%%%%%%%%%%%%%%%%%%%%%%%%%%%%%%%%%%%%%%%%%%

%\bibliographystyle{prsty_4etal}
\bibliographystyle{prsty}
\input epsf
\def\ns2{NbSe$_2$}

\title {Comment on  `On the Luttinger theorem concerning the number of particles in the
ground states of systems of interacting fermions', arXiv:0711.0952v1, by B. Farid}

\author{ A. Rosch}

\affiliation{Institute for Theoretical Physics, University of Cologne, 50937 Cologne, Germany}
\date{\today}

\begin{abstract}
  In his preprint \cite{farid}, arXiv:0711.0952v1, Behnam Farid argues
  that the Luttinger theorem is valid not only for a metal but also
  for a Mott insulator if the chemical potential is calculated by
  taking the limit of vanishing temperature at fixed particle density.
  In contrast, we have found in our recent paper \cite{rosch} on the
  basis of a controlled strong coupling expansion that the Luttinger
  theorem is violated in this limit for a particle-hole asymmetric
  two-band Mott insulator. In an extensive discussion of our result
  Farid argues that an arbitrarily weak breaking of particle-hole
  symmetry leads to a destruction of the Mott insulating state at half
  filling. In this comment we point out that this is not correct.
%
%In our recent paper \cite{rosch}
% we have shown using a controlled strong coupling
%expansion that this is not the case
%
%
%In his preprint \cite{farid}, arXiv:0711.0952v1, Behnam Farid claims
%that in our recent paper \cite{rosch} we used the limit of vanishing temperature, $T \to 0$, in a wrong way, when
%discussing the breakdown of the Luttinger theorem for a two-band Mott insulator.
%We argue that this is not the case and point out that
\end{abstract}
%\pacs{}
\maketitle
\newcommand{\w}{\omega}

 \acknowledgments

In a recent paper \cite{rosch}, we have shown
that in a certain two-band Hubbard model in its Mott insulating phase the Luttinger theorem
(in the variant given below) is not valid
for a range of chemical potentials.
The Luttinger theorem
\begin{eqnarray}\label{luttinger}
n &=& 2 \sum_\alpha \int_{G_\alpha(\bf{p},\omega=0)>0} \frac{d^3 \vec{p}}{(2 \pi)^3}
\end{eqnarray}
relates the density of particles, $n$,  to a volume in momentum space where the Greens function at $\w=0$ is
 positive. Here
$\alpha$ is a band index.
In the case of a Mott insulator, the Greens function changes its sign without
 having a pole at the so-called Luttinger surface, see references in \cite{farid} and \cite{rosch}.

It is very easy \cite{rosch} to convince oneself that the right-hand
side of Eq. (\ref{luttinger}) depends on the chemical
potential $\mu$ when $\mu$ is varied {\em within} the gap of a Mott insulator [as $G({\bf p},\w,\mu)=G({\bf p},\w+\mu)$].
In contrast, the
left-hand side is independent of $\mu$ at $T=0$ for any value of $\mu$ within the gap.
Therefore
there is typically only a single value of the chemical potential, $\mu=\mu_L$, within the gap
where  (\ref{luttinger}) is valid (as shown explicitely in Ref.~\cite{rosch}).

Farid \cite{farid} agrees that the Luttinger theorem is not valid for a range of chemical potentials but
argues that no such problem can arise when the chemical potential is calculated
in the limit $T \to 0$ at fixed particle density $n$, $\mu_n=\lim_{T \to 0} \mu(n,T)$ (Farid uses the notation
$\mu_\infty=\mu_n$).

In contrast, we obtained in Ref. \cite{rosch} within a controlled
strong coupling expansion that even in this limit, the Luttinger
theorem is not valid  for a generic particle-hole asymmetric situation, $\mu_n \neq \mu_L$. We used that $\mu_n$ is located
in the middle of the gap such that the activation energies of many-particle eigenstates
with particle number $N-1$ and $N+1$ are identical. A calculation of
$\mu_n$ (and $\mu_L$) to linear order in $1/U$ turns out to be
sufficient \cite{rosch} to construct a counter example to the Luttinger theorem for
$\mu=\mu_n$.

On the pages 58 to 86 of his preprint Farid discusses these questions,
analyzes our result and comes to the conclusion that we determined
$\mu_n$ in an incorrect way when taking the limit of zero temperature
at fixed particle density. His argument is based on a surprising
result of his calculations: he claims \cite{cite} that an arbitrarily
small breaking of particle-hole symmetry transforms the half-filled
Mott insulator into a metal, or, equivalently, that the particle-hole
asymmetric system is not half-filled if the chemical potential is
located within the gap!  In our opinion, this is obviously wrong. For
example, it contradicts the observation that small perturbations have
no effects in systems with a finite gap (in the two-band Mott
insulator under consideration both the charge and the spin gap are
finite).  In the appendix \ref{appendix} we sketch the formal argument
which can be used to prove this.

 But it is also useful to check how the surprising result is obtained in Ref.~\cite{farid}
 that the Mott insulating state is not half filled.
In Eq. (6.50) the electronic density, $n$, of the Mott insulating
phase is calculated by using formula (6.36) [or, equivalently (6.43)] for the Greens function
which is obviously only valid up to order $t/U$. In the absence of
particle-hole symmetry, Farid argues that the deviation from half-filling, $n-1$, is finite at $T=0$
and of order $(t/U)^3$ (see sentence below Eq. (6.59)). It is, however, not possible to calculate a quantity
to order $(t/U)^3$ based on an approximate formula which is valid only up to order $t/U$.

The wrong interpretation of this calculation seems to be
the  reason, why Farid disagrees with our finding  \cite{rosch} that the Luttinger
theorem is not  valid for
$\mu=\mu_n$. 

I would like to thank B. Farid for discussions which helped me to understand his point of view.

%\begin{appendix}
\section*{Appendix: Particle density of a Mott insulator}\label{appendix}
In this appendix we briefly sketch a formal argument which can be used to proof that a Mott insulator remains
 half-filled at $T=0$ in the presence of weak particle-hole asymmetry. Here we consider the
 two-band Mott insulator defined in Ref. \cite{rosch} which has a unique ground state and a
 gap not only in the charge sector but also in the spin sector.
 We start with a particle-hole symmetric model with $N_0$ sites. Due to the particle-hole symmetry,
 the system is half-filled, $N=\langle \hat{N} \rangle =N_0$. Now, we track the evolution of the ground state
when particle-hole symmetry is broken, e.g. by switching on a weak next-nearest neighbor hopping $t'$.
As $[H,\hat{N}]=0$, the ground state is always an eigenstate of $\hat{N}$, and as $\hat{N}$ has a discrete spectrum,
the groundstate remains exactly half-filled as long as there is no level-crossing.
As the gap of the Mott insulator is finite in the thermodynamic limit, no such level crossing can occur for small
$t'$.

%\end{appendix}
%We acknowledge useful discussions with M. Vojta.

\end{document}